\documentclass[acmsmall,nonacm]{acmart}

\usepackage{enumitem}
\usepackage{amsmath}
\usepackage{booktabs}
\usepackage{tabularx}
\usepackage{array}

\title{Characterizing Experiments with Synthetic MSI/HSI Data: A Structured Taxonomy for Agri-Food Research}

\author{Jasmine Battestin Nunes}
\affiliation{%
  \institution{Universidade Estadual de Campinas (UNICAMP)}
  \city{Campinas}
  \country{Brazil}
}

\author{Paula Dornhofer Paro Costa}
\authornote{Corresponding author.}
\affiliation{%
  \institution{Universidade Estadual de Campinas (UNICAMP)}
  \city{Campinas}
  \country{Brazil}
}
\email{paulad@unicamp.br}

\begin{document}

\begin{abstract}
Synthetic multispectral and hyperspectral data are increasingly used to address limited measurements, labels, and acquisition resources in spectral-imaging research. Yet the scope of the evidence produced by these experiments depends not only on the synthesis method, but also on sensing conditions, sample variability, data provenance, evaluation design, and downstream use. We present a literature-informed taxonomy for experiments using synthetic MSI/HSI data through spectral reconstruction and/or data augmentation, with a deliberate focus on agri-food research. The taxonomy is paired with a traversal procedure that connects available evidence, sensing constraints, experimental choices, and evaluation design to the forms of generalization that were actually tested, producing a compact experiment characterization record. Although primarily intended for studies with an existing research question, the same structure can also support earlier-stage planning when only an object of study, dataset, or sensing setup is initially defined. The framework was developed through question-driven reading, practical research experience, and iterative refinement against additional recent studies, and is offered as a basis for further use and refinement.
\end{abstract}

\maketitle

\section{Introduction}

Multispectral imaging (MSI) and hyperspectral imaging (HSI) combine spatial and spectral information and are widely used in agri-food research to non-destructively assess composition, quality, maturity, defects, contamination, and related physical or chemical attributes. However, acquisition cost and complexity, limited availability of diverse datasets, and reference measurements that may be destructive, slow, labor-intensive, or expensive constrain their use~\cite{Amigo2019SettingScene,GraciaMoises2023DA,Ahmed2025Review}.

Synthetic spectral data provide two complementary routes for working under these constraints. Spectral reconstruction estimates richer spectral information from lower-dimensional or indirect measurements, while Data Augmentation (DA) expands or diversifies the evidence available for model development. Both have been explored in agri-food settings, including synthetic NIR generation, generative spectral augmentation, and RGB-to-HSI reconstruction~\cite{Sa2022DeepNIR,Zhang2022MaizeDCGAN,Zhao2020Tomato}.

The challenge motivating this work is not the heterogeneity of these experiments itself. Rather, the conditions that delimit their evidence are often distributed across descriptions of sensors, samples, preprocessing, model design, evaluation, and downstream tasks. As a result, it can be difficult to determine which sources of variation were represented, which remained fixed, and which forms of transfer were actually tested.

We therefore propose a taxonomy for experiments involving synthetic MSI/HSI data through spectral reconstruction and/or DA, with an explicit focus on agri-food applications. Its first role is descriptive: to organize the physical, experimental, computational, and task-related choices that define a study. Its second role is procedural: by traversing the taxonomy, a researcher can connect the available evidence, sensing constraints, and evaluation design to an \emph{experiment characterization record} that makes the corresponding generalization boundaries easier to inspect. While this is primarily intended for studies with an existing research question, the same structured traversal can also support earlier-stage planning when only an object of study, dataset, sensing resource, or other part of the experiment is initially defined. Sections~\ref{sec:taxonomy} and~\ref{sec:using} present these two central components. The preceding sections provide only the context, scope, and development rationale needed to interpret them; a terminology appendix is provided for readers from neighboring disciplines. The taxonomy is conceived as an evolving framework rather than a fixed specification, and perspectives from the broader research community are welcomed as part of its continued refinement.

\section{Related Work}
\label{sec:background}

The proposed structure draws on three connected bodies of work. First, spectral-reconstruction literature shows that the problem formulation depends strongly on what is measured, what is reconstructed, the sensing model, and the evaluation target~\cite{Arad2016Sparse,Zhang2022Survey,Ahmed2025Review}. Second, DA literature in optical spectroscopy and agri-food applications emphasizes limited or imbalanced data, the distinction between simple perturbations and learned generation, and the need to relate augmentation to the intended learning task~\cite{GraciaMoises2023DA,Zhang2022MaizeDCGAN}. Third, application studies often evaluate reconstructed, synthetic, or fused spectral information through a downstream objective, such as tomato-quality estimation, fruit detection, or food-quality prediction~\cite{Zhao2020Tomato,Sa2022DeepNIR,Li2023Fusion}. Together, these works motivate treating sensing, synthesis, downstream use, and evaluation as parts of the same experimental description.

The documentation perspective of Gebru et al.'s \emph{Datasheets for Datasets} also provides useful methodological inspiration~\cite{Gebru2021Datasheets}. Their proposal uses structured questions and a workflow to document dataset motivation, composition, collection, use, and related considerations, with the broader aim of supporting reflection and informed use. We adopt an analogous principle --- that structured prompts can make assumptions and omissions easier to see --- but apply it to a narrower object: experiments involving synthetic spectral data. The resulting record is therefore complementary to, rather than a substitute for, dataset documentation.

\section{Scope}
\label{sec:scope}

The taxonomy targets experiments in which synthetic MSI/HSI information is produced or used through \textbf{spectral reconstruction}, \textbf{data augmentation}, or both. It covers cases in which the synthetic spectral product is evaluated and cases in which it serves as an intermediate representation for prediction, detection, segmentation, grading, visualization, or decision support.

The focus on agri-food is deliberate. The literature studied for this work, and much of the practical context that motivated it, concerns biological variability across specimens, batches, cultivars, maturity or storage stages, defects, and acquisition conditions, together with costly or destructive reference measurements. The taxonomy is therefore not presented as already validated for MSI/HSI research in general. Although it was developed for agri-food, researchers may test it in other MSI/HSI domains. Such use would provide an external check of its scope by showing which categories remain useful, which require adaptation, and which domain-specific factors are missing.

\section{Taxonomy Development}
\label{sec:development}

The taxonomy was developed through a question-driven study of the literature rather than a formal systematic review. We studied foundational HSI concepts from the general spectral-imaging literature~\cite{Amigo2019SettingScene}. The literature study then focused on two topics particularly relevant to the taxonomy: computational spectral reconstruction~\cite{Zhang2022Survey,Ahmed2025Review} and data augmentation in agri-food spectroscopy~\cite{GraciaMoises2023DA}. Categories were added or reorganized when recurring experimental questions could not be represented clearly by the existing structure.

This literature-based process was complemented by practical research experience with spectral-imaging experiments and reconstruction tasks. An initial version of the taxonomy was then compared 
against additional papers from the 2026 IEEE International Conference on Acoustics, Speech and Signal Processing (ICASSP) as an additional refinement step. We used LLM-assisted screening to identify potential missing branches, ambiguous terminology, and cases that did not fit the existing structure. These cases were subsequently inspected manually and used to revise the taxonomy where appropriate.

\section{The Taxonomy}
\label{sec:taxonomy}

The taxonomy contains four macro-parts. \textbf{A. Reconstruction} describes how spectral information is acquired, reconstructed, fused, and evaluated. \textbf{B. Data Augmentation} describes why and how synthetic or transformed data expand training evidence. \textbf{C. Downstream tasks in agri-food} records the practical role for which spectral information is used. \textbf{D. Research opportunities} collects recurring gaps exposed by the first three branches. A--C primarily characterize the experiment itself; D is more naturally used after characterization to articulate unresolved limitations or future directions. Section~\ref{sec:using} turns these branches into a decision-oriented traversal whose main purpose is to connect experimental characterization and evaluation to the resulting generalization boundary. The same traversal can also begin earlier, when only an object of study, dataset, or sensing setup is available, because the detailed characterization reveals which experimental decisions and evidence are still missing.

\subsection*{A. Reconstruction}

This section explores the computational recovery of hyperspectral and multispectral data from simpler, lower-dimensional acquisitions. At first glance, the task of spectral reconstruction --- such as mapping three broadband visible channels (RGB) to dozens of narrow spectral bands --- is severely underconstrained: different spectral distributions can produce the same or very similar RGB responses, a phenomenon known as \emph{metamerism}~\cite{Arad2016Sparse,Zhang2022Survey}. A direct one-to-one analytical inversion from RGB to a full spectrum is therefore generally unavailable without additional assumptions or learned priors.

The feasibility of data-driven reconstruction relies on regularities in the spectra encountered in the target domain. Arad and Ben-Shahar, for example, motivate RGB-to-HSI recovery by assuming that relevant natural spectra occupy a comparatively low-dimensional subset of the space of all possible spectra, making informative priors possible~\cite{Arad2016Sparse}. In agri-food experiments, the validity of such priors depends on how well the training data represent the products, acquisition conditions, and sensing system the model will be used on.

An important boundary concerns extrapolation beyond the spectral regime represented by the input measurements and training data. Signoroni et al. reported that the RGB-to-HSI deep-learning approaches covered by their 2019 review were limited to the visible spectrum and noted that, at that time, they were not aware of DL-based MSI-to-HSI spectral upsampling in the NIR--SWIR range (750--3000 nm)~\cite{Signoroni2019Review}. This historical observation should not be read as a claim that longer-wavelength properties can never be statistically predicted from RGB. Rather, when the input device does not directly sense target wavelengths, the prediction necessarily depends on correlations learned from paired data and should be evaluated as such. Experimental claims should therefore distinguish between reconstructing information supported by the sensing configuration and predicting spectral regions that are not directly measured by the input.

To navigate this complex process, the taxonomy in this section is structured as a continuous experimental pipeline. It begins by establishing the input-output architecture (\textbf{A1. Problem setting}) and the fundamental light-matter interactions at play (\textbf{A2. Physical basis}). Subsequently, it maps the practical boundaries and environmental conditions of the acquisition setup (\textbf{A3. Experimental factors and sensing constraints}). With the physical problem contextualized, the taxonomy categorizes the algorithmic engines used for inversion (\textbf{A4. Reconstruction methods}) and the frameworks for combining multiple data sources (\textbf{A5. Fusion strategies}). Finally, it outlines an assessment structure to quantify the physical fidelity and practical reliability of the generated spectral cubes (\textbf{A6. Evaluation}).

\subsubsection*{A1. Problem setting}
First, it is important to clearly define the \textbf{problem setting}, i.e., the high-level input-output structure of the reconstruction problem. At this stage, the goal is to state what type of measurement enters the pipeline and what spectral product should come out. A project may involve a single mapping, such as $\text{RGB}\rightarrow\text{HSI}$, or a chain of stages, such as $\text{Mosaic}\rightarrow\text{RGB}\rightarrow\text{HSI}$.
\begin{enumerate}[label=\roman*)]
    \item \textbf{RGB $\rightarrow$ HSI:} reconstruct a hyperspectral image from a conventional RGB image. This is the standard spectral reconstruction setting when deployment relies on commodity color cameras, but the task benefits from dense spectra.
    
    \item \textbf{RGB $\rightarrow$ NIR:} estimate near-infrared information from RGB data. This is appropriate when the downstream signal of interest lies mainly in the NIR range, but only RGB sensing is available at test time. It can be one single NIR channel (for instance, generating synthetic NIR images to improve agricultural fruit detection systems~\cite{Sa2022DeepNIR}) or several of them (the latter being a subcase of the RGB $\rightarrow$ MSI or HSI setting).

    \item \textbf{RGB $\rightarrow$ MSI (Targeted Reconstruction):} estimate a specific, discrete set of task-relevant multispectral bands from a standard RGB input. This bypasses the need to reconstruct a full continuous spectrum when only a few chemical anchors (e.g., water or chlorophyll peaks) are required for the agricultural task.
    \item \textbf{MSI $\rightarrow$ HSI (Spectral Super-Resolution):} reconstruct a continuous, dense hyperspectral cube from a limited set of discrete multispectral bands. This is highly relevant when test-time deployment uses commercially available, low-cost multispectral cameras, but downstream processing requires fine spectral resolution.
    \item \textbf{Multi-sensor spectral fusion:} combine complementary/related measurements from two or more sensors, such as RGB, NIR, MSI, HSI, depth, or thermal data, to obtain a richer spectral product. This setting is useful when no single sensor is sufficient on its own, but the measurements are related and can be fused; or when different sensors provide complementary or redundant information that can improve reconstruction, prediction, robustness, or interpretability compared with using a single source alone.
    \item \textbf{Mosaic $\rightarrow$ HSI:} reconstruct a full hyperspectral cube from mosaiced spectral measurements, typically produced by a spectral filter array. Mosaiced spectral measurements are closer to raw acquisition than RGB, because the input already contains band-specific samples, but not a full spectrum at every pixel.
    \item \textbf{Indirect acquisition $\rightarrow$ HSI:} reconstruct an HSI from measurements that do not directly provide one spatial image per spectral band.
    
    \begin{enumerate}[label=\alph*)]
        \item \textbf{Snapshot spectral imaging $\rightarrow$ HSI:} recover or organize an HSI from a spectral measurement captured in a single exposure. This setting is mainly defined by acquisition time: the scene is captured at once, rather than by scanning points, lines, or wavelengths. It is useful for moving samples, real-time inspection, or unstable scenes, but it often requires computational steps to separate, rearrange, or estimate the spatial--spectral information recorded by the sensor.
        \item \textbf{Compressive spectral imaging $\rightarrow$ HSI:} reconstruct an HSI from coded or multiplexed measurements in which spatial and spectral information are intentionally mixed before reaching the sensor. This setting is mainly defined by the measurement model: the recorded data are not a direct spectral cube, so reconstruction depends on knowing how the scene was encoded and on using priors or learned models to recover the missing spatial-spectral information. A compressive system may be snapshot or multi-shot.
        \item \textbf{Single-pixel spectral imaging $\rightarrow$ HSI:} reconstruct an HSI from a sequence of coded measurements acquired with a non-imaging detector. Instead of recording a spatial image directly on a pixelated sensor, the scene is measured through known illumination or modulation patterns, and each measurement contains a mixed response from the scene. Spatial and spectral information are then recovered computationally from the pattern sequence. This setting is useful when pixelated spectral sensors are unavailable, expensive, noisy, or difficult to use in the target wavelength range.
    \end{enumerate}
    
    \item \textbf{MSI/HSI restoration / deconvolution:} start from an already acquired MSI or HSI and recover a cleaner or more reliable version of the same spectral cube. \emph{Restoration} is the broader task, including denoising, destriping, artifact correction, and quality improvement. \emph{Deconvolution} is a specific restoration case focused on reversing blur or point-spread effects introduced by optics, motion, focus, or the acquisition system. The task is not to infer spectra from another modality, but to improve a spectral cube that has already been measured.
    
    \item \textbf{MSI/HSI super-resolution / pansharpening:} increase the spatial resolution of an MSI or HSI while preserving its spectral information. \emph{Super-resolution} is the broader task: it may use only the low-resolution cube or may use an additional high-resolution guide image. \emph{Pansharpening} is a specific fusion-based case, originally defined as the fusion of a multispectral (MS) image with a panchromatic (PAN) image to combine the spectral information of the MS image with the spatial information of the PAN image \cite{Amigo2019Pansharpening}; extended to modern hyperspectral imaging, the goal is similarly to inject high spatial detail into the spectral cube without introducing spectral distortion.
    
\end{enumerate}

\subsubsection*{A2. Physical basis}
After defining the high-level problem setting, it is essential to establish the \textbf{physical basis} of the measurements. Unlike standard computer vision tasks, hyperspectral and multispectral imaging in agri-food rely heavily on precise light-matter interactions that reveal intrinsic chemical, biological, and physical properties of the samples. This section maps the physical reality of the data before treating it as purely mathematical arrays, a critical step for robust experimental design.

\begin{enumerate}[label=\roman*)]
    \item \textbf{Image formation and sensing model:} defines the ``journey of the photon'' --- how light interacts with the sample or object of study and is ultimately converted into a digital signal. In the present agri-food scope, that object is typically a food or agricultural product. Understanding this process is key to designing valid reconstruction methods.
    \begin{itemize}
        \item \emph{What is measured?} The physical interaction between light and the sample. For instance, in a fruit, diffuse reflectance includes light that penetrates the skin, interacts with cellular structures, and scatters back toward the sensor, carrying information related to the material's composition and structure.
        \item \emph{How is it measured?} The acquisition mechanism capturing the scene, such as a push-broom scanner moving over apples on a conveyor belt, or a snapshot camera over a packing line.
        \item \emph{How is it encoded?} The data format generated by the sensor, such as a raw 2D mosaic from a spectral filter array or discrete multiplexed bands.
        \item \emph{What is lost or mixed?} The physical limitations of the capture. This includes limited light penetration depth into the food's tissue, spatial-spectral mixing in compressive sensors, or spectral blurring due to overlapping filter sensitivities.
    \end{itemize}

    \item \textbf{Reflectance vs. raw values:} distinguishes the raw camera response from a signal corrected to better represent the sample's spectral response. A \emph{raw value} (often a Digital Number - DN) is the immediate sensor output and depends jointly on illumination, sensor sensitivity, acquisition settings, and the optical properties of the sample. \emph{Reflectance} expresses the fraction of incident light reflected by the sample as a function of wavelength, typically after reference correction. The combined influence of source, sample, and sensor is one reason preprocessing and calibration are central in MSI/HSI analysis~\cite{AmigoSantos2019Preprocessing}.

    \item \textbf{Radiometric and inter-device calibration:} the physical and mathematical procedures used to convert raw sensor outputs into a calibrated or relative reflectance representation. To reduce the influence of illumination and sensor dark current, spectral-imaging systems commonly use a white reference image ($W$, representing a high-reflectance reference under the acquisition illumination) and a dark reference image ($D$, representing the sensor response without incident light). A common relative-reflectance calculation for the raw target image $I$ is:
    $$R = \frac{I - D}{W - D}$$
    Reflectance calibration and illumination-related calibration problems are discussed specifically by Geladi et al.~\cite{Geladi2004Calibration}. Documenting the calibration procedure is important for interpreting spectral values and for assessing comparability across devices, acquisition sessions, or environments.

    \item \textbf{Relevant wavelength regions:} relates the measured spectral range to biological, chemical, or physical properties of the sample or object of study. In agri-food applications, visible and infrared regions can carry information associated with pigments, water, proteins, structural changes, and other constituents or properties. The interpretation of spectral regions depends on the material, acquisition mode, and task:
    \begin{itemize}
        \item \emph{Visible (VIS):} commonly contains information related to color, pigments, and surface appearance.
        \item \emph{NIR and SWIR:} can contain overtone and combination-band responses associated with molecular bonds such as O--H, C--H, and N--H, and are widely used in compositional and quality analyses.
    \end{itemize}
    These general associations are part of the broader MSI/HSI background summarized in the book edited by Amigo~\cite{Amigo2019HyperspectralImaging}. In a specific experiment, however, establish task-relevant wavelengths from the material, sensing configuration, and empirical evidence rather than assuming a generic spectral region.
\end{enumerate}

\subsubsection*{A3. Experimental factors and sensing constraints}

While the physical basis (A2) defines the interaction between light, sample, and sensor, section \textbf{A3} maps the practical choices and sources of variation in the experimental setup. Changes between controlled laboratory conditions and deployment environments can alter the data distribution and model performance. These eight blocks are therefore used to define dataset boundaries, acquisition constraints, and preprocessing decisions. Depending on the research question, they can also identify variables to be varied explicitly in evaluation or represented through physically plausible data-augmentation procedures.

\begin{enumerate}[label=\roman*)]
    \item \textbf{Camera / sensor variability:} addresses the hardware dependency of the acquired signal. Different camera models --- or even different units of the same model --- have unique spectral response curves, quantum efficiencies, and noise profiles. It is crucial to state whether the method is strictly tied to a single sensor or whether cross-camera robustness is evaluated.
    
    \item \textbf{Data representation:} defines the digital format in which the signal enters the computational pipeline. This includes the stage of the camera pipeline (e.g., raw sensor data, demosaiced images, or ISP-processed outputs), the input type (e.g., RGB, sparse MSI, compressive measurements), and data preservation formats (e.g., bit depth and lossless vs. lossy compression). These factors dictate the theoretical maximum fidelity a reconstruction model can achieve.
    
    \item \textbf{Spectral definition:} outlines the specific spectral information available for the task. It involves documenting the total number of channels, the covered spectral range (e.g., 400--1000 nm), and the sensor band design. The latter is a critical physical parameter that dictates the width of each spectral band (Full Width at Half Maximum - FWHM) and the degree of overlap between adjacent bands. In targeted applications, justifying the band selection (informative vs. redundant bands) is crucial for building lightweight and deployment-ready models.
    
    \item \textbf{Acquisition conditions:} catalogs the physical capture environment, which heavily biases the data distribution. A comprehensive methodology requires documenting the illumination (light source, intensity, and spectral profile), the acquisition geometry (working distance and spatial resolution), and the collection environment (e.g., controlled lab vs. dynamic field). Furthermore, dynamic scene factors like background clutter, object occlusion, and sample motion (e.g., moving on a conveyor belt) should be considered as key variables for evaluating model robustness and generalizability under physical perturbations.
    
    \item \textbf{Product / sample coverage:} represents the biological variability included in the dataset. Agri-food products can vary across species or cultivars, batches or lots, maturity or storage stages, regions, seasons, defects, and spoilage states. Limited coverage of such factors can restrict the forms of biological generalization that the experiment is able to evaluate.
    
    \item \textbf{Pre-processing choices:} encompasses deterministic transformations applied before modeling. These can include radiometric corrections, region-of-interest (ROI) isolation, smoothing, normalization, derivatives, Standard Normal Variate (SNV), and Multiplicative Scatter Correction (MSC). Such operations can reduce noise or scattering-related variation, but they also modify the representation seen by the model and should therefore be documented and evaluated for the specific dataset~\cite{AmigoSantos2019Preprocessing}.
    
    \item \textbf{Alignment between sensing streams:} critical for multi-sensor setups and multi-modal fusion. Spatial and temporal registration constraints need to be addressed. If an RGB camera and an MSI sensor have different fields of view or trigger times, the resulting pixel misalignment must be computationally corrected or explicitly handled by a misalignment-robust reconstruction model.
    
    \item \textbf{Available acquisition metadata:} details the supplementary instrumental information preserved alongside the dataset, such as calibration files, spectral response curves, integration times, sensor gain, and white/dark reference measurements. Preserving this information can support interpretation, recalibration, simulation, inter-device comparison, and later domain-adaptation studies.
\end{enumerate}

\subsubsection*{A4. Reconstruction methods}
Once the physical constraints (A2) and real-world variabilities (A3) are mapped, the next step is selecting the core algorithmic ``logic'' or architecture used to solve the inverse problem of spectral reconstruction. Hyperspectral and multispectral imaging systems combine conventional imaging and spectroscopy, essentially mapping spatial and spectral dimensions simultaneously. Consequently, extracting reliable signatures from the resulting massive datasets requires robust computational approaches. 

In agri-food research, method choice depends partly on the amount and diversity of paired data available and on the conditions under which the model is expected to operate. A direct data-driven mapping can fit a controlled paired dataset well yet degrade when illumination, sensor response, or sample distribution changes. Prior-based or physics-informed constraints may be useful when paired data are limited or when explicit structure is important.

Ablation studies and cross-family baseline comparisons can help determine which components contribute to an observed gain, without treating such comparisons as proof that one architectural choice is universally superior. Zhang et al. organize computational spectral reconstruction around a fundamental distinction between prior-based and data-driven approaches~\cite{Zhang2022Survey}. Building on that distinction, this taxonomy uses four practical families:

\begin{enumerate}[label=\roman*)]
    \item \textbf{Prior-based:} these approaches constrain the solution using explicit assumptions or structure, such as sparsity, low-dimensional representations, smoothness, or statistical/physical models. Classical chemometric methods can also provide useful constrained or low-dimensional baselines. For example, PCA represents data through orthogonal directions of decreasing explained variance, while MCR formulations may impose constraints such as non-negativity or closure when those constraints are appropriate to the problem.

    \item \textbf{Direct data-driven methods:} these approaches learn a mapping from the input measurement to the target spectral representation from paired examples. This family includes CNN/residual/U-Net/dense-based networks and attention/transformer-based models. Their flexibility can model nonlinear relationships without requiring a fully specified analytical inverse model, but performance remains dependent on the coverage and representativeness of the training data.

    \item \textbf{Cross-modal generative methods:} these methods formulate reconstruction as conditional generation or image-to-image translation between sensing representations. GAN-based or related generative models may combine reconstruction losses with adversarial or perceptual objectives. Pretrained visual representations can sometimes be reused through transfer learning, but their usefulness depends on the target modality and spectral task. The same generative family may also appear in Section B when its role is to create additional training data rather than reconstruct a paired target.

    \item \textbf{Physics-informed methods:} these approaches incorporate known physical structure into a learned reconstruction process. The broader physics-informed machine-learning literature describes ways of combining data with mathematical or physical models, including architectures with built-in physical structure and objectives or constraints derived from governing knowledge~\cite{Karniadakis2021Physics}. In spectral reconstruction, this can include:
    \begin{itemize}
        \item \emph{Physics-based data or simulation:} forward models are used to generate or perturb training observations under controlled assumptions.
        \item \emph{Physics-aware architecture:} the model structure reflects an acquisition or optimization process, for example through model-based or unfolded networks.
        \item \emph{Physics-aware objectives or constraints:} camera response functions, forward models, mixing assumptions, or other domain knowledge are incorporated into the optimization objective or output constraints.
    \end{itemize}
    Such approaches can improve physical consistency and may improve transfer in some settings, but those benefits still need to be established empirically for the target experiment.
    
\end{enumerate}

\subsubsection*{A5. Fusion strategies}
Researchers increasingly rely on multi-sensor setups to evaluate complex agricultural and food products. Rather than just combining sensors when a single one is insufficient, fusion strategies are actively used when different sensors provide complementary or redundant information that can significantly improve reconstruction, prediction, robustness, or interpretability compared with using a single source alone. However, merely concatenating datasets does not guarantee better performance; it can introduce noise and curse-of-dimensionality issues. This section maps the strategies used to merge multimodal information. For a rigorous experimental design, any proposed fusion model typically involves strict ablation studies, to prove that the fusion genuinely yields advantages. Furthermore, a robust approach theoretically and empirically justify \emph{how} and \emph{where} the modalities are merged using the following frameworks:

\begin{enumerate}[label=\roman*)]
    \item \textbf{By model stage:} this categorization stems primarily from deep learning and neural network architectures and describes where information from different modalities is combined within the model.
    \begin{itemize}
        \item \emph{Early fusion:} modalities are concatenated at the input level before passing through the network layers. \emph{Example:} Stacking a high-resolution RGB image and a registered low-resolution HSI cube into a single multi-channel input tensor before feeding it into a Convolutional Neural Network (CNN) for fruit quality grading.
        \item \emph{Late fusion:} each modality is processed by independent neural branches (sub-networks), and their learned feature representations or final prediction scores are merged only at later layers. This is highly effective when the modalities represent fundamentally different physical phenomena. \emph{Example:} Using a 2D-CNN to extract spatial features from an RGB image and a 1D-CNN to extract chemical features from HSI spectra, concatenating them right before the final classification layer to predict ripeness.
    \end{itemize}

    \item \textbf{By representation level:} heavily rooted in classical data fusion and chemometrics theory, this paradigm dictates the level of data abstraction prior to the fusion.
    \begin{itemize}
        \item \emph{Low-level fusion (LLF):} integrates measurements from multiple sources before feature extraction. \emph{Example:} Li et al. generated LLF data by directly concatenating Vis-NIR and NIR spectral variables for chicken-quality prediction~\cite{Li2023Fusion}. When LLF is performed pixel-by-pixel across separate imaging systems, spatial or temporal registration may additionally become an important experimental constraint.
        \item \emph{Intermediate-level fusion (ILF):} also known as feature-level fusion, this strategy extracts representative variables from each source before merging them. \emph{Example:} features or selected wavelengths from different sensors can be combined into a reduced representation before prediction. In the chicken study of Li et al., the best LLF model was obtained for total viable count (TVC), whereas the best result for total volatile basic nitrogen (TVB-N) was obtained with ILF~\cite{Li2023Fusion}. This illustrates that the preferred fusion level can depend on the target variable.
        \item \emph{High-level fusion (HLF):} also known as decision-level fusion, it builds independent predictive models for each data source and combines their outputs, for example by averaging probabilities or using a voting rule. Because interaction occurs only after the modality-specific predictions have been formed, this strategy does not directly model cross-modal relationships at the raw-data or intermediate-feature levels.
    \end{itemize}

    \item \textbf{Misalignment-robust fusion:} separate sensors may produce spatially or temporally misaligned observations because of differing fields of view, frame rates, motion, or parallax. Early or pixel-level fusion can be sensitive to these mismatches. Misalignment-robust strategies therefore attempt to combine information without assuming exact correspondence, for example through learned alignment, attention, deformable operators, or feature-level matching. Studies using multi-sensor data should state whether correspondence is assumed, estimated, or handled explicitly.
\end{enumerate}

\subsubsection*{A6. Evaluation}
The final stage of the experimental pipeline is evaluating the performance of the proposed methods. In the context of hyperspectral imaging for agri-food, evaluation is notably complex: a model that produces visually pleasing images or achieves a low average mathematical error might completely destroy the specific, narrow chemical absorption bands required for food quality prediction. A robust evaluation framework extends beyond standard computer vision metrics, employing a holistic approach that measures physical fidelity, downstream usefulness, and industrial reliability. A comprehensive evaluation framework can be structured around these four pillars:

\begin{enumerate}[label=\roman*)]
    \item \textbf{Reconstruction metrics:} quantify agreement between reconstructed and measured spectral data. MRAE, RMSE, and SAM are three metrics commonly used in RGB-to-HSI reconstruction; Zhang et al. review these metrics and their limitations, and Zhao et al. use all three in an agri-food reconstruction study~\cite{Zhang2022Survey,Zhao2020Tomato}. They provide complementary views:
    \begin{itemize}
        \item \emph{Magnitude error:} RMSE and MRAE summarize numerical differences between reconstructed and reference values using different normalizations.
        \item \emph{Spectral shape:} SAM measures the angle between reconstructed and reference spectral vectors and is therefore sensitive to spectral-direction differences rather than only absolute magnitude.
        \item \emph{Spatial or application-specific fidelity:} when the task depends on spatial structure, color, or another domain-specific property, additional metrics such as SSIM, PSNR, or task-specific measures can complement spectral errors.
    \end{itemize}
    No single metric should be interpreted as a complete description of reconstruction quality.

    \item \textbf{Downstream-task validation:} measures the practical utility of the reconstructed data. While reconstructed spectral cubes can be the final product (e.g., when expanding databases for Data Augmentation), they frequently serve as intermediate inputs for predicting specific traits. In such cases, evaluating the downstream-task performance provides a highly relevant validation path. A comprehensive evaluation often seeks to demonstrate that the reconstructed HSI/MSI performs comparably to ground-truth data in practical applications, such as fruit detection, quality parameter estimation, or spoilage assessment. For example, spatial assessment is an important tool for microbial characterization; specifically, a reconstruction model might achieve a low numerical error but blur the image texture features required for the early detection of bacterial biofilms on various materials. A model might exhibit a generic absolute error but excel in downstream validation if it successfully preserves the specific task-relevant wavelengths.

    \item \textbf{Interpretability / uncertainty:} complementary analyses can examine whether the reconstructed signal behaves consistently with the intended physical or downstream use. Depending on the study, this may include sensitivity or feature-attribution analyses, comparison of reconstructed spectral curves or derivatives with measured references, inspection of latent or multivariate structure, or uncertainty estimates. These analyses should be treated as additional evidence rather than substitutes for reconstruction and downstream evaluation.

    \item \textbf{Robustness / transfer evaluation:} tests selected boundaries of the model's generalization capabilities and links evaluation back to the factors mapped in A3. When a study makes a transfer claim, the corresponding source of change should be represented in the evaluation design, for example through cross-camera testing, changed acquisition conditions, or held-out batches or harvest periods. The tested axis should be stated explicitly rather than summarized as generic ``robustness.''
\end{enumerate}

\subsection*{B. Data Augmentation}

Data augmentation broadens the effective training distribution without requiring every additional observation to be independently acquired and annotated. This is particularly relevant in agri-food spectroscopy, where data collection may be constrained by sampling periods, environmental conditions, class imbalance, and the cost or effort of obtaining sufficiently representative measurements~\cite{GraciaMoises2023DA}. Within this taxonomy, DA is treated broadly: it includes direct perturbation or resampling of measured data, learned generative synthesis, and strategies that expand supervision when labels or paired measurements are scarce.

A central distinction is that augmentation is not synonymous with reconstruction. Reconstruction estimates or transforms a measured representation, for example RGB $\rightarrow$ HSI. Augmentation changes the amount or diversity of training evidence. The two may nevertheless be combined: reconstructed spectra can become synthetic training samples, while augmentation can increase the effective training set available to a reconstruction model.

\subsubsection*{B1. Why augment}
B1 records the experimental constraint that motivates augmentation. The same technique can be appropriate for one constraint and inappropriate for another, so the motivation should be stated before the mechanism.
\begin{enumerate}[label=\roman*)]
    \item \textbf{Small datasets:} augmentation increases the number or diversity of training observations when measured data are limited. It does not, however, create new independent biological specimens by itself.
    \item \textbf{Class imbalance:} targeted augmentation can increase representation of rare defects, contamination states, quality grades, or other minority conditions during training.
    \item \textbf{Stabilize training:} controlled perturbations may act as regularization and reduce sensitivity to incidental properties of a small training set.
    \item \textbf{Improve generalization:} augmentation may deliberately expose a model to plausible changes in intensity, noise, geometry, or spectral response. Such perturbations are most interpretable when linked to physical or acquisition variability identified in A3.
    \item \textbf{Paired-data scarcity:} reconstruction pipelines may require aligned measurements from two sensors or modalities. Synthetic pairing, unpaired translation, or related strategies can reduce dependence on perfectly registered pairs, although synthetic targets should not be treated as measured ground truth.
    \item \textbf{Label scarcity:} chemical references, microbiological measurements, masks, and expert annotations can be costly. Augmentation can increase the effective use of labeled observations, while B5 covers methods that expand supervision more directly.
\end{enumerate}

\subsubsection*{B2. Uses of DA}
B2 records where augmented or synthetic observations enter the experimental pipeline.
\begin{enumerate}[label=\roman*)]
    \item \textbf{Support reconstruction training:} transformations expand the training distribution of a reconstruction model. When correspondence is required, paired modalities should normally receive consistent spatial transformations.
    \item \textbf{Support downstream-task training:} augmented MSI/HSI, spectra, patches, or targets are used to train the final classifier, regressor, detector, or segmenter. The maize-kernel study of Zhang et al. is one example of generative augmentation supporting a quantitative downstream task~\cite{Zhang2022MaizeDCGAN}.
    \item \textbf{Pretraining / transfer:} synthetic or augmented data are used to learn an initial representation that is subsequently adapted to a target dataset or task. The experiment should distinguish clearly which stages use real and synthetic observations.
    \item \textbf{Support cross-scene adaptation:} augmentation is designed to bridge an identified domain shift, such as laboratory-to-field changes, new backgrounds, or illumination differences. Evidence for this purpose requires an evaluation in the target condition rather than only a random split of the source condition.
\end{enumerate}

\subsubsection*{B3. Non-DL augmentation}
This branch contains deterministic or stochastic transformations that do not require a deep generative model.
\begin{enumerate}[label=\roman*)]
    \item \textbf{Cropping:} extracts spatial subregions or patches. It increases the number of training instances but not the number of independent specimens.
    \item \textbf{Oversampling:} repeats or preferentially resamples observations, commonly to rebalance classes. It changes the training distribution without adding new measured information.
    \item \textbf{Noise injection:} adds perturbations intended to mimic detector noise or measurement variability. The noise model should be justified when robustness conclusions depend on it.
    \item \textbf{Spectral shift / intensity variation:} perturbs wavelength position, scale, offset, or intensity. Excessive or physically implausible changes can move or distort chemically meaningful bands.
    \item \textbf{Blend spectra:} combines spectra or image regions to synthesize intermediate observations. The mixing rule should be consistent with the physical assumptions of the application.
    \item \textbf{SMOTE:} interpolates between neighboring observations in a feature space. These samples are mathematical constructions and should not automatically be interpreted as physically realizable spectra.
\end{enumerate}

\subsubsection*{B4. Generative augmentation}
Generative augmentation learns a distribution from observed data and samples new spectra, images, or paired modalities from that learned distribution. GAN-based approaches, including DCGAN and semi-supervised GAN variants, have been used for spectral augmentation in agri-food studies~\cite{GraciaMoises2023DA,Zhang2022MaizeDCGAN}. The same type of model can appear in different parts of the taxonomy depending on what it does in the experiment: for example, a GAN belongs to A4 when it maps one sensing modality to another for reconstruction, but to B4 when it is used to create additional training samples. Reporting should include what is generated, what conditions the generator, which real data were used to fit it, and whether generated samples are filtered or validated before use.

\subsubsection*{B5. Supervision expansion}
B5 captures methods whose principal purpose is to increase usable supervision.
\begin{enumerate}[label=\roman*)]
    \item \textbf{Pseudo-labeling:} a model assigns labels or continuous targets to previously unlabeled observations. Reference labels and pseudo-labels should remain distinguishable in the experiment record.
    \item \textbf{Synthetic paired targets:} artificial targets are generated so that more input--target pairs become available. Such targets inherit assumptions and errors from the mechanism that created them.
    \item \textbf{Self-supervised surrogate targets:} supervision is derived from the data themselves through masking, reconstruction, contrastive objectives, or other proxy tasks before task-specific adaptation.
\end{enumerate}

\subsubsection*{B6. Synthetic data sources}
The final DA branch records the provenance of synthetic observations.
\begin{enumerate}[label=\roman*)]
    \item \textbf{Learned generation:} data are generated by a model fitted to examples, such as a GAN, diffusion model, autoencoder, translator, or reconstruction network. Its support is bounded by the training data and model assumptions.
    \item \textbf{Physical simulation:} data are generated through an explicit model of illumination, material response, sensor sensitivity, filtering, noise, geometry, or another part of the acquisition process. Its credibility depends on the adequacy and parameterization of the forward model.
\end{enumerate}
Hybrid pipelines should record both sources rather than collapsing them into a generic ``synthetic'' label.

\subsection*{C. Downstream Tasks in Agri-Food}

Section C records what the spectral information is ultimately used for. This branch matters because the meaning of ``good'' synthetic data is task dependent. A reconstruction can have low average spectral error and still fail to preserve a narrow feature needed for chemical prediction; conversely, an imperfect reconstruction may remain useful for a particular decision task. C should therefore be read together with A6.

\subsubsection*{C1. Detection / localization / counting}
This category covers the presence, location, or number of fruits, food products, defects, foreign materials, or other target objects. Spectral information can be the primary signal or an auxiliary modality. Synthetic NIR for fruit detection is one example~\cite{Sa2022DeepNIR}. The experiment record should state the localization unit and whether synthetic data are used during training, inference, or both.

\subsubsection*{C2. Quality attribute estimation}
This category covers continuous or ordinal product properties such as composition, maturity, moisture, oil content, soluble solids content, titratable acidity, firmness, or related quality indicators. Evaluation should identify how the reference target was measured and at what experimental unit the train/test split was performed. Reconstructed HSI for tomato-quality estimation and augmented NIR-HSI for maize oil prediction illustrate this type of task~\cite{Zhao2020Tomato,Zhang2022MaizeDCGAN}.

\subsubsection*{C3. Safety / spoilage / defect assessment}
This branch includes contamination, microbial spoilage, anomalies, damage, adulteration, and related integrity or safety indicators. The record should distinguish a directly measured property from a statistical proxy associated with it, and should state the reference method used to define the target.

\subsubsection*{C4. Segmentation / grading / sorting}
This category groups tasks that delineate regions, assign products to quality categories, or route products according to a decision rule. Although all may use classification internally, they differ in output granularity and deployment requirements. The experiment should therefore specify whether the unit of prediction is a pixel, region, specimen, batch, or process decision.

\subsubsection*{C5. Visualization / decision support}
Spectral information may support human interpretation through chemical maps, quality maps, process monitoring, or harvest/storage decisions. Visual plausibility alone is not evidence of physical validity; substantive interpretation should remain connected to a measured quantity, validated spectral feature, or explicitly defined decision process.

\subsection*{D. Research Opportunities}

Unlike A--C, Section D is not primarily a description of what the experiment is. It is a diagnostic branch that collects recurring gaps made visible by characterization. A study may relate to one or more D categories because it attempts to address a documented limitation, but it is not expected to instantiate every branch.

\subsubsection*{D1. Collect more data}
This opportunity concerns insufficient coverage of the physical or biological domain. Relevant directions include paired RGB--MSI/HSI/NIR acquisition, labeled downstream-task datasets, temporal or longitudinal measurements, broader object/scene/condition coverage, standardized acquisition metadata, and validation of pseudo-labels or synthetic targets. The emphasis is on independent coverage rather than raw sample count.

\subsubsection*{D2. Learn compatible representations}
Different sensors may vary in band count, wavelength range, response functions, calibration, spatial resolution, or measurement principle. Opportunities include band-agnostic or channel-flexible architectures, spectral resampling and band mapping, cross-camera training, cross-sensor domain mapping, calibration transfer, and representations that bridge RGB, mosaic, compressive, and single-pixel inputs.

\subsubsection*{D3. Improve generalization}
This branch targets failure under unseen products, batches, conditions, devices, scenes, or imperfect multimodal alignment. Relevant directions include broader object coverage, cross-condition and cross-device benchmarks, alignment-aware models, and adaptation with limited target supervision. ``Generalization'' should always be qualified by the axis along which the domain changes.

\subsubsection*{D4. Make objectives more useful}
Reconstruction metrics and practical utility are not equivalent. Research opportunities include studying when fidelity metrics predict downstream performance, developing task-aware losses, optimizing joint fidelity--utility objectives, and defining robustness metrics that explicitly quantify stability across sensors or acquisition conditions.

\subsubsection*{D5. Combine DA and reconstruction}
DA and reconstruction can be studied jointly: augmentation may compensate for limited paired supervision in reconstruction, while reconstruction models may generate spectral samples for downstream augmentation. Such pipelines require clear provenance tracking and careful train/test separation so that synthetic descendants of evaluation samples do not enter training.

\subsubsection*{D6. Physics-informed pipelines}
This branch concerns stronger integration of sensing physics and light--matter interaction into synthetic-data pipelines. Physics-informed machine learning broadly combines data-driven models with physical models, constraints, or physically structured architectures~\cite{Karniadakis2021Physics}. In MSI/HSI pipelines, corresponding directions include physics-informed simulation, hybrid physics-based and learned models, physical constraints on outputs, and use of measured sensor response functions for cross-device mapping. These approaches make assumptions more explicit but do not remove the need for empirical validation.

\section{Using the Taxonomy to Characterize Experiments and Generalization Boundaries}
\label{sec:using}

The primary use considered here is a researcher who already has a scientific question or practical task. In this case, the taxonomy helps connect the available evidence, experimental choices, and evaluation design to the forms of generalization that the experiment can actually support. The research problem itself does not need to be reformulated simply to use the taxonomy.

The same structure can also be entered at an earlier stage. A researcher may have only an object of study, a set of samples, an existing dataset, an available sensing setup, or an interest in applying MSI/HSI to a domain. In that case, the taxonomy can be used prospectively to organize what is already known, identify plausible downstream questions and sensing requirements, and expose what would still need to be defined or measured before a specific experiment can be established. This is a secondary use of the framework: the detailed characterization required by the taxonomy can also support problem formulation and experiment planning.

In both cases, the traversal follows the same underlying logic: record what is already fixed, characterize the available evidence and sensing resources, identify what is missing or differs across sources, define the role of synthetic spectral information, specify the experiment and its evaluation, and finally derive the generalization boundary supported by the resulting evidence. It is not necessary to populate every taxonomy branch; only the branches relevant to the study and to the interpretation of its evidence need to be recorded.

\subsection{Decision-oriented traversal}

Table~\ref{tab:record} summarizes the procedure. The steps are sequential for clarity but can be revisited as the study develops. For example, identifying an important evidence gap in Step~3 may lead to new data collection, after which the evidence inventory should be reconsidered. The final column links each step to the relevant taxonomy branches and to the concrete result that should be available after that step.

\begin{table}[!htbp]
\caption{Decision-oriented traversal of the taxonomy.}
\label{tab:record}
\centering
\footnotesize
\setlength{\tabcolsep}{1.8pt}
\renewcommand{\arraystretch}{0.90}
\begin{tabularx}{\linewidth}{
    >{\centering\arraybackslash}p{0.024\linewidth}
    >{\raggedright\arraybackslash}p{0.155\linewidth}
    X
    >{\raggedright\arraybackslash}p{0.19\linewidth}}
\toprule
\textbf{\#} &
\textbf{Question} &
\textbf{What the researcher does} &
\textbf{Taxonomy / result} \\
\midrule

1 &
What is already defined? &
If the research question is established, record the target product or population, task, intended use, and expected sensing conditions. If it is not, record the available object, samples, dataset, or apparatus and use the taxonomy to formulate a candidate experimental target. &
A1--A2 + C $\rightarrow$ starting point and target statement. \\

2 &
What evidence and sensing resources are available? &
Characterize relevant datasets and the available apparatus: independent samples, references, sensors, spectral coverage, band configuration, products or cultivars, batches, acquisition conditions, preprocessing, alignment, and metadata. &
A2--A3 + C $\rightarrow$ evidence and sensing inventory. \\

3 &
What is missing or differs across sources? &
Compare the intended experiment with the available datasets and apparatus. Identify missing samples, references, modalities or conditions, as well as sensor or acquisition mismatches, and determine whether existing data, transfer, calibration or band mapping, or new acquisition can address them. &
A2--A3 $\rightarrow$ evidence-gap and acquisition/transfer plan. \\

4 &
What role should synthetic data play? &
Determine whether the study uses reconstruction, DA, or both, and record why. If multiple sensing sources are involved, determine whether fusion is also part of the pathway. &
A1, A4--A5 + B1--B2 $\rightarrow$ synthetic-data role and rationale. \\

5 &
How will the experiment be implemented and evaluated? &
Specify acquisition and preprocessing, reconstruction/fusion/augmentation methods, supervision and synthetic-data provenance, experimental unit, train/validation/test split, reconstruction and/or downstream evaluation, and the factors relevant to the intended transfer claim. &
A2--A6 + B3--B6 + C $\rightarrow$ experiment specification and evaluation plan. \\

6 &
What does the experiment actually support? &
Compare the conditions represented during development and evaluation. Record relevant factors as varied, held fixed, or not observed/not applicable; identify which transfer axes were directly tested; then use D to organize the remaining limitations and opportunities. &
A3 + A6, then D $\rightarrow$ generalization boundary and experiment characterization record. \\

\bottomrule
\end{tabularx}
\end{table}

Whether the available evidence is adequate in Step~3 depends on the intended experiment rather than on a universal minimum sample count. A dataset containing thousands of images may still represent only a small number of independent specimens, batches, or acquisition conditions. Conversely, a smaller dataset may be adequate for a deliberately narrow question. Adequacy should therefore be assessed from both the number of independent experimental units and the biological, physical, and acquisition variability required by the intended evaluation.

Consider, for example, a researcher whose problem is already defined: estimating soluble-solids content in melons. The researcher can begin directly from that target and examine whether suitable melon datasets already exist. Through A2--A3, the available evidence is characterized in terms of independent fruits, cultivars, batches, sensors, wavelength coverage, acquisition conditions, and reference measurements. If paired RGB--HSI melon data are available, RGB $\rightarrow$ HSI reconstruction can be investigated directly; if the paired data are limited, DA may additionally be considered. If such a dataset is not available, an RGB--HSI dataset from another crop, such as tomatoes, may instead motivate a transfer experiment: for example, testing whether a reconstruction model learned on tomatoes remains useful when applied to melons. The evaluation then depends on the measurements available for the melon experiment. Measured melon HSI allows reconstructed and measured spectra to be compared, whereas a laboratory reference such as soluble-solids content allows the researcher to test whether the reconstructed representation is useful for that downstream quality task.

The same example can begin earlier. A researcher may simply have access to melon samples and become interested in MSI/HSI without yet having a specific research question. Section~C can help organize candidate downstream purposes --- for example, quality estimation, defect assessment, or sorting --- while A1--A2 help connect those purposes to possible sensing or reconstruction settings. The researcher then proceeds to the same evidence inventory and evidence-gap analysis before defining the remainder of the experiment.

An available experimental setup can itself be part of this starting point. Suppose the researcher has melon samples and access to a particular RGB, MSI, HSI, or multi-sensor system and intends to collect a new dataset. A2--A3 make the constraints of that setup explicit: sensor response, spectral range and band configuration, spatial resolution, illumination, geometry, calibration, preprocessing, alignment, and available metadata all delimit what the new dataset can contain. These characteristics can then be compared with existing datasets. Similar configurations may facilitate reuse or combination, whereas differences in sensors, bands, calibration, or acquisition conditions may motivate resampling, calibration transfer, domain adaptation, or a deliberate cross-device or cross-condition experiment. Once the dataset is collected, those same apparatus choices become part of its characterization and later define which forms of generalization can and cannot be evaluated.

Thus, whether the researcher begins from a finished question, an existing dataset, a collection of samples, or an available apparatus, the taxonomy eventually brings the study to the same point: an explicit experimental target, a documented evidence base, and a clear account of how synthetic spectral information will be produced and evaluated.

\subsection{From experimental design to a generalization boundary}
\label{sec:generalization}

The central interpretive step is to connect the factors recorded in A3 with the evaluation recorded in A6. Generalization should be reported along explicit axes rather than treated as a single property of a model. For each factor relevant to the intended use, the record should indicate which values or conditions were represented during model development and which were represented during evaluation.

A factor being \emph{varied} is not, by itself, evidence of transfer. Suppose three cultivars occur throughout both training and test data. The experiment contains cultivar variability, but it has not tested performance on an unseen cultivar. By contrast, training on cultivars A and B and evaluating on cultivar C directly tests transfer to cultivar C. The same distinction applies to batches, seasons, illumination conditions, cameras, acquisition environments, and other factors.

Three situations are therefore useful to distinguish. If a factor remains the same throughout development and evaluation, it is held fixed and transfer along that axis is not tested. If several levels are represented but all are encountered during model development, the experiment evaluates performance within the represented variation. If evaluation contains a level or condition absent from model development, the experiment directly tests transfer along that axis.

The unit of the train/validation/test split is part of this interpretation. MSI/HSI experiments may generate many pixels, patches, spectra, repeated views, paired measurements, or augmented descendants from the same physical specimen. Their numerical abundance does not make them independent replicates. Treating dependent observations as independent is related to pseudoreplication, and validation procedures should account for relevant grouped or hierarchical structure~\cite{Hurlbert1984Pseudoreplication,Roberts2017CrossValidation}.

Accordingly, the split should be defined at the level corresponding to the intended claim before derived observations are generated. If the claim concerns unseen specimens, specimens should be separated before extracting patches. If it concerns unseen batches, batches should define the split. If it concerns cross-device transfer, at least one device must be absent from model development and reserved for evaluation.

Synthetic or transformed descendants of an observation should likewise remain on the same side of the train/test boundary as their source. Otherwise, information from an evaluation unit may influence model development and yield an overoptimistic performance estimate, a form of data leakage~\cite{KapoorNarayanan2023Leakage}.

At the end of the traversal, the \emph{experiment characterization record} should make six elements explicit:

\begin{itemize}[leftmargin=*,nosep]
    \item the study starting point and experimental target;
    \item the available evidence and the physical and acquisition scope of the measured data;
    \item the identified evidence gaps and the reason for new acquisition, transfer, reconstruction, DA, fusion, or another strategy;
    \item the measured and synthetic data pathway, including preprocessing, provenance, and supervision;
    \item the evaluation design, including the experimental unit, split, references, metrics, and downstream assessment; and
    \item the variation map, directly tested transfer axes, resulting generalization boundary, and remaining gaps.
\end{itemize}

The purpose of the record is not to maximize the number of taxonomy branches represented. It is to make explicit the chain connecting the study's starting point, available evidence and sensing constraints, experimental choices, evaluation, and the generalization boundary supported by the resulting experiment.

\section{Conclusion}

We propose a literature-informed taxonomy and traversal procedure for characterizing experiments that use synthetic MSI/HSI data through spectral reconstruction and/or data augmentation, with a deliberate focus on agri-food research. Its central purpose is to connect experimental design and evaluation to the scope of the evidence produced, making explicit which forms of generalization were actually tested. The proposal does not attempt to certify robustness, replace dataset documentation, or define a universal standard.

By recording the available evidence and sensing constraints together with the experimental choices, and by distinguishing which relevant factors were varied, held fixed, or left unobserved, the taxonomy provides a route from technical description to a structured experiment characterization record. The same detailed traversal can also support earlier-stage research when only an object of study, dataset, sensing setup, or other part of the experiment is initially defined, by exposing which questions, measurements, and experimental decisions still need to be established. We view the present taxonomy as a starting point: useful insofar as researchers can apply it, identify where it fails, and improve it.

\bibliographystyle{ACM-Reference-Format}
\bibliography{refs}

\clearpage
\appendix
\section{Terminology for Interdisciplinary Readers}
\label{sec:terms}

This appendix provides concise working definitions for terms that recur throughout the paper and may be less familiar to readers from neighboring disciplines. The definitions are intended to support interpretation of the taxonomy rather than to provide exhaustive or normative definitions.

For readers seeking broader background, three references provide useful entry points. The book \emph{Hyperspectral Imaging}, edited by Jos\'e Manuel Amigo, provides broad coverage of MSI/HSI foundations, preprocessing, data analysis, and applications~\cite{Amigo2019HyperspectralImaging}. Zhang et al.~\cite{Zhang2022Survey} provide a focused overview of computational spectral reconstruction from RGB images, including image formation, method families, datasets, and evaluation. For data augmentation in the agri-food context, Gracia Mois\'es et al.~\cite{GraciaMoises2023DA} review augmentation strategies for optical spectroscopy, from simple transformations to generative approaches. Additional references are included below only where a term involves a methodological issue that is particularly relevant to interpreting an experiment.

\begin{description}

    \item[RGB, MSI, and HSI.]
    RGB images contain three broad color channels. Multispectral imaging (MSI) records measurements in a limited set of spectral bands, while hyperspectral imaging (HSI) generally samples the spectrum more densely and produces many neighboring bands. An HSI is commonly organized as a data cube whose axes describe two spatial coordinates and wavelength. The distinction between MSI and HSI is not determined by a universal numerical threshold; band number, spacing, width, and spectral coverage depend on the sensor and application.

    \item[Spectral band and wavelength range.]
    A \emph{spectral band} represents the response measured over a wavelength interval. The \emph{wavelength range} is the overall spectral interval covered by a sensing system. Labels such as VIS, NIR, VNIR, and SWIR denote regions of the electromagnetic spectrum, although their exact boundaries may vary between fields and instruments.

    \item[Raw sensor values, reflectance, and calibration.]
    A raw sensor value, often represented as a digital number (DN), is the numerical response recorded by the acquisition system. It depends not only on the sample but also on factors such as illumination, detector response, exposure settings, and sensor noise. Reflectance measurements aim to express the fraction of incident radiation reflected by the sample under specified measurement conditions. Calibration procedures, such as white and dark referencing in reflectance imaging, are used to reduce acquisition-system effects and obtain a representation that is more comparable across measurements.

    \item[Preprocessing.]
    Preprocessing refers to transformations applied to measured data before subsequent modeling or analysis. Depending on the experiment, these may include calibration, smoothing, normalization, derivatives, correction of scattering effects, or spatial operations such as region-of-interest selection. Preprocessing transforms observations that already exist; data augmentation, in contrast, deliberately creates additional or modified training instances.

    \item[Paired data and registration.]
    \emph{Paired data} are corresponding measurements of the same sample or scene in two representations or modalities, for example an RGB image and the HSI used as its reconstruction target. \emph{Registration} refers to aligning those measurements spatially and, when relevant, temporally. Imperfect registration can make nominally paired observations inconsistent at the pixel or region level.

    \item[Spectral reconstruction, inverse problem, and metamerism.]
    \emph{Spectral reconstruction} estimates a richer spectral representation from another measurement, such as RGB $\rightarrow$ HSI or MSI $\rightarrow$ HSI. Such problems can be \emph{ill posed}: the available measurement may not determine a unique spectral solution without additional assumptions or information. In RGB-based reconstruction, one reason is \emph{metamerism}, whereby different spectral distributions can produce the same or very similar RGB responses. Zhang et al.~\cite{Zhang2022Survey} provide a broader treatment of computational RGB-to-HSI reconstruction and its underlying image-formation problem.

    \item[Data augmentation (DA).]
    Data augmentation increases or diversifies the training examples available to a model through transformations, resampling, synthetic generation, or related strategies. Importantly, creating several augmented versions of one measured specimen increases the number of training instances but does not create additional independently collected specimens. Gracia Mois\'es et al.~\cite{GraciaMoises2023DA} provide a review focused specifically on augmentation for optical spectroscopy in agri-food applications.

    \item[Supervision, pseudo-labeling, and self-supervised learning.]
    In \emph{supervised learning}, examples are associated with reference labels or target values used during training. In \emph{pseudo-labeling}, predictions made by a model are used as provisional targets for otherwise unlabeled observations. In \emph{self-supervised learning}, the training signal is constructed from the data themselves through a surrogate task, after which the learned representation may be adapted to the task of interest. These strategies differ in where the information used as supervision originates.

    \item[Training and inference.]
    \emph{Training} is the stage in which model parameters are estimated from development data. \emph{Inference} is the subsequent use of the fitted model to generate an output for a new input. This distinction is important in multimodal experiments because a sensor or modality used to construct supervision during training does not necessarily need to be available at inference time.

    \item[CNN, GAN, and transformer.]
    These terms denote broad families of neural-network architectures. 

    \item[Feature or latent representation.]
    A feature representation is a numerical description used by a model instead of, or in addition to, the original measurements. A \emph{latent representation} usually refers to an internal representation learned by a model. In multimodal fusion, data can therefore be combined at the level of raw measurements, intermediate representations, or final predictions.

    \item[Transfer learning and domain adaptation.]
    \emph{Transfer learning} refers broadly to reusing knowledge learned in one setting to support learning in another. \emph{Domain adaptation} is a related setting in which the source and target data differ in their distributions and the method seeks to reduce or accommodate that difference. In this paper, these terms are relevant when models, representations, or calibration knowledge are transferred across datasets, sensors, products, or acquisition conditions.

    \item[Ground truth and reference measurement.]
    These terms denote the observation against which a prediction or reconstruction is evaluated. Depending on the task, the reference may be a measured HSI cube, a laboratory chemical analysis, a microbiological measurement, or a manually annotated image. The term \emph{reference measurement} is useful in experimental settings because the reference itself may also be subject to measurement uncertainty.

    \item[Benchmark.]
    A benchmark combines data and an evaluation protocol to provide a common basis for comparing methods. Performance on a benchmark describes performance under the conditions represented by that benchmark and protocol; it does not, by itself, establish performance under sensors, samples, or acquisition conditions that were not evaluated.

    \item[Loss function and evaluation metric.]
    A \emph{loss function} is a quantity optimized, directly or indirectly, while fitting a model. An \emph{evaluation metric} is a quantity used to assess its resulting performance. The two can be the same, but they need not be. For example, a reconstruction model may be optimized using one loss while being evaluated using several complementary spectral, spatial, and downstream metrics.

    \item[Train/validation/test split and experimental unit.]
    Training data are used to fit a model, validation data to make development choices such as hyperparameter or model selection, and test data to estimate performance after those choices have been made. The \emph{experimental unit} is the independently sampled entity relevant to the scientific question, such as a specimen, batch, acquisition session, or sensor. When observations are grouped or otherwise dependent, splitting individual observations at random can place closely related measurements in both development and evaluation sets.

    \item[Data leakage.]
    Data leakage occurs when information that should belong only to evaluation influences model development, whether through training examples, preprocessing, feature construction, model selection, or another part of the pipeline. Leakage can produce overoptimistic performance estimates; Kapoor and Narayanan provide a detailed taxonomy of leakage in machine-learning-based science~\cite{KapoorNarayanan2023Leakage}. In synthetic-data experiments, a particularly relevant case occurs when an original sample contributes related or augmented observations to both training and test data.

    \item[Pseudoreplication.]
    In this paper, \emph{pseudoreplication} is used in the broad experimental-design sense of treating dependent measurements as though they were independent replicates. Thus, thousands of pixels or patches extracted from a small number of specimens do not become thousands of independently collected specimens. Hurlbert's classical treatment defines the problem in the context of treatment effects when treatments are unreplicated or replicates are not statistically independent~\cite{Hurlbert1984Pseudoreplication}.

    \item[Domain shift, generalization, and robustness.]
    A \emph{domain shift} occurs when relevant data-generating conditions differ between model development and evaluation or use, for example because the sensor, batch, cultivar, illumination, or environment changes. \emph{Generalization} concerns performance on observations not used to fit the model, whereas \emph{robustness} concerns stability under a specified source of variation or perturbation. In this paper, these terms are most informative when the corresponding axis of change is stated explicitly.

    \item[Downstream task.]
    A downstream task is the practical task that uses the spectral information produced or processed earlier in the pipeline, such as detection, quality prediction, spoilage assessment, segmentation, grading, or visualization. Downstream evaluation therefore asks whether synthetic spectral information is useful for its intended purpose, rather than only whether it numerically resembles a reference spectrum.

    \item[Baseline and ablation study.]
    A \emph{baseline} is a reference method or experimental configuration against which another method is compared. An \emph{ablation study} removes, replaces, or varies a component or design choice while keeping the remainder of the experiment as comparable as possible. Its purpose is to investigate which components contribute to the observed result.

    \item[Physics-informed machine learning.]
    Physics-informed machine learning incorporates known physical structure into a data-driven modeling process. Physical knowledge can enter through simulated observations, model architecture, constraints, forward models, or learning objectives. The broader relationship between physical modeling and machine learning is reviewed by Karniadakis et al.~\cite{Karniadakis2021Physics}. In the present taxonomy, the relevant physics primarily concerns sensing and light--matter interaction.

    \item[Curse of dimensionality.]
    In this paper, the \emph{curse of dimensionality} refers to difficulties that arise when a high-dimensional representation must be learned from limited independent data. As dimensionality grows, more observations are generally needed to estimate statistical structure and decision boundaries reliably. This issue is particularly relevant to HSI, where each observation may contain tens or hundreds of spectral bands while the number of independently collected labeled samples may be comparatively small; Landgrebe discusses hyperspectral analysis explicitly as a high-dimensional signal-processing problem~\cite{Landgrebe2002Hyperspectral}.

\end{description}

\end{document}